\documentclass[a4paper,10pt,conference]{style/IEEEtran}

\usepackage{amsmath,amsfonts,amssymb}
\usepackage{graphicx,subfigure}
\usepackage[ruled]{algorithm}
\usepackage{algpseudocode}
\usepackage{float}
\usepackage{graphicx}
\usepackage{flushend}
\usepackage{placeins}
\usepackage{xcolor}
\usepackage{soul}
\newcounter{example}
\newtheorem{exampleaux}[example]{Example}
\newenvironment{example}[1][]
    {\vspace*{.2em}\begin{exampleaux}\textit{#1}
    \normalfont}
    {\hfill $\square$\end{exampleaux}}

\begin{document}

\title{\LARGE \bf
Active search and coverage using point-cloud reinforcement learning
\vskip -0.5em
}

\author{\IEEEauthorblockN{Matthias Rosynski}
\IEEEauthorblockA{Department of Automation\\Technical University of Cluj-Napoca\\Cluj-Napoca, Romania\\Email: rosynski.matthias@aut.utcluj.ro}
\and
\IEEEauthorblockN{Alexandru Pop}
\IEEEauthorblockA{Department of Automation\\Technical University of Cluj-Napoca\\Cluj-Napoca, Romania\\Email: alexandru.pop@aut.utcluj.ro}
\and
\IEEEauthorblockN{Lucian Bu\c soniu}
\IEEEauthorblockA{Department of Automation\\Technical University of Cluj-Napoca\\Cluj-Napoca, Romania\\Email: lucian.busoniu@aut.utcluj.ro}

}
\maketitle
\thispagestyle{empty}
\pagestyle{empty}

\makeatletter
\def\blfootnote{\gdef\@thefnmark{}\@footnotetext}
\makeatother
\blfootnote{\textbf{Acknowledgement:} This work was financially supported from H2020 SeaClear, a project that received funding from the European Union’s Horizon 2020 research and innovation programme under grant agreement No. 871295, and by the project “Network of excellence in applied research and innovation for doctoral and postdoctoral programs / InoHubDoc”, co-funded by the European Social Fund financing agreement no. POCU/993/6/13/153437. Computation resources were provided by the CLOUDUT Project, co-founded by the European Fund of Regional  Development through the Competitiveness Operational Programme 2014-2020, contract no. 235/2020.}

\begin{abstract}

We consider a problem in which the trajectory of a mobile 3D sensor must be optimized so that certain objects are both found in the overall scene and covered by the point cloud, as fast as possible. This problem is called target search and coverage, and the paper provides an end-to-end deep reinforcement learning (RL) solution to solve it. The deep neural network combines four components: deep hierarchical feature learning occurs in the first stage, followed by multi-head transformers in the second, max-pooling and merging with bypassed information to preserve spatial relationships in the third, and a distributional dueling network in the last stage. To evaluate the method, a simulator is developed where cylinders must be found by a Kinect sensor. A network architecture study shows that deep hierarchical feature learning works for RL and that by using farthest point sampling (FPS) we can reduce the amount of points and achieve not only a reduction of the network size but also  better results. We also show that multi-head attention for point-clouds helps to learn the agent faster but converges to the same outcome. Finally, we compare RL using the best network with a greedy baseline that maximizes immediate rewards and requires for that purpose an oracle that predicts the next observation. We decided  RL achieves significantly better and more robust results than the greedy strategy.

\end{abstract}

\section{Introduction}

Searching for target objects in an environment is an important goal of active robotic sensing. These methods are crucial for search and rescue operations, but can also be employed in other applications of autonomous aerial, underwater, or surface vehicles.
We present in this paper an end-to-end reinforcement learning (RL) approach for target search and coverage based on a 3D point cloud representation of the scene. A point cloud is a discrete set of points in space represented by their Cartesian coordinates. The agent learns an optimal strategy not only to find the target objects but also to cover them, increasing the likelihood e.g. that the target objects can subsequently be  correctly classified  \cite{krishna2020adaptive}. 

Deep learning is a machine learning method based on neural networks, which are particularly computationally expensive when going from a 2D pixel representation to a 3D voxel form. Scaling up the resolution is prohibitive since the compute and memory costs of voxel-based models grow cubically with input resolution. We can reduce the input space by modifying the representation, because many voxels do not contain any information. For this reason, the point cloud representation is one of the most widely used in the field of 3D object classification and segmentation.
Our framework also relies on point clouds, which describe a canonical space: the observations can be projected in a standard form and do not rely on non-essential environmental features, unlike images and many other representations. This is essential when transitioning from simulation to real-world applications \cite{james2019sim,lobos2020point}. 

Contrary to  2D image processing, point clouds present a challenge for today's neural networks. It is difficult to apply convolutional layers that take advantage of spatial or serial dependence due to the fact that point clouds lack a discernible order. Qi et al. \cite{qi2017pointnet} proposed PointNet, a convolution architecture with a max-pooling layer to make the function symmetric and hence independent of the order of the points. Several enhancements were made to this architecture, the most notable being PointNet++ \cite{qi2017pointnet++}, which captures local structures caused by the metric space. ShellNet \cite{zhang2019shellnet} focuses on decreasing the network's complexity in order to shorten the training period. With the advent of transformers, which were introduced for translation models \cite{vaswani2017attention}, a new possibility appeared to implement neural networks for point clouds. The most well-known approach is the Point Cloud Transformer (PCT) network \cite{guo2021pct}, which learns weight matrices to represent the correlations between the individual points.

In RL, an agent learns from experience gained from interacting with its environment. Each experience sample consists of a state, an action performed by the agent in that state, a reward 
that serves as feedback for the agent, and the following state. These samples are stored in off-policy algorithms \cite{sutton2018reinforcement} in a so-called experience replay buffer, from which the neural network is trained. The objective is to maximize the cumulative, long-term rewards.

Inspired by the PCT network and a few other methods, we devise here a novel RL method that exploits a new network structure to adress target search and coverage. The network combines four distinct stages: a modified embedding stage, the multihead attention stage, a max pooling stage, and the RL stage which includes the distributional dueling network architecture \cite{bellemare2017distributional, dueling}. 

We specified three classes in the scene: wall, floor and target object, which must be distinguishable in advance, and which indicate distinct exploration priorities. 
We implemented an RL agent and compared it with a greedy baseline. 
The greedy baseline selects an action that maximizes the reward for a single step.
The reward for both methods is the difference between the amount of points before and after taking a measurement. Since the greedy approach must (unlike RL) predict rewards, it requires an oracle that can foresee point cloud measurements.

\subsection*{Related work}
To the best of our knowledge, point-cloud-based RL for target search and coverage problems has not been the subject of any research. 
Nonetheless, object coverage for autonomous inspection tasks is an active research subject, in which an agent finds the most efficient global path to cover an object. Landgraf et al. \cite{landgraf2021reinforcement} devised a view path planning technique that uses RL to cover an object. They turned a point cloud into a CAD model, which along with the agent's position forms the state. The reward was the information gain over an unseen area. The action space is continuous and the selected action is the next pose / viewpoint of the agent, from where the next snapshot will be taken which can be anywhere in the environment. After selecting the next pose, the agent employs a second algorithm to calculate the path to it. In contrast to this approach, our RL methodology takes point clouds directly as agent states. In addition, our objective differs in that we are not just covering an object, but also searching for it in its surroundings.
Furthermore, the actions are different, as our agent can only move one step relative to his current location and it cannot ``teleport'' across the scene.

A recent RL study demonstrates how to use directly a point cloud as input to solve a navigation problem \cite{lobos2020point}. In that paper, a network architecture based on ShellNet is proposed. ShellNet calculates features regarding the vicinity of certain representative points without encoding the representative points themselves, making it inapplicable for RL. Therefore, the authors of \cite{zhang2019shellnet} altered the network structure to preserve this geometric information.
Due to the fact that pooling is performed on each ``ShellConv'' layer before the layers are combined, spatial information will still be lost.
By exploiting the segmentation technique from the PointNet \cite{qi2017pointnet} network, we chose to utilize a different strategy to preserve geometrical information. In addition, we do  active coverage as well.

The point-cloud-based RL study \cite{krishna2020adaptive} investigated the problem of finding a query object with unknown position and pose within a given point cloud. We attempt to incrementally construct in the most efficient way a point cloud that encompasses the target objects, whereas \cite{krishna2020adaptive} uses the complete point cloud from the start and seeks to identify the object within it. 

Research was also undertaken in the field of point cloud-based RL for sim-to-real dexterous manipulation \cite{qin2022dexpoint}, in order for a robot to perform specific tasks, like grasping objects or opening doors, through a pointNet-based network. Nevertheless, this application differs from our work.
 
Reference \cite{tiator2020point} segments a point cloud using RL  and a PointNet-based network, whereas we employ a framework that combines PCT classification, PointNet segmentation, and RL rainbow network, to which we also added multi-head attention.

We demonstrate that, contrary to recent techniques in the field of two-dimensional active target sensing using RL \cite{goel2022reinforcement} or informative path planning \cite{popovic2020informative}, not only is it possible to learn how to search for target objects in a three-dimensional environment, but also to cover them, directly from three-dimensional shape information. The more information about the object the data structure, the more features the network can recognize. This can be useful later e.g. for object inspection or classification.

\section{Problem Description}
Consider a point cloud $S$, which is defined as a set of points $\left\{P_i \in  \mathbb{R}^{3}\mid i=1,...,n \right\}$, and represents a scene containing objects of interest. The point cloud is collected incrementally by a moving sensor, described by its position and orientation $q_t \in \mathcal{R}^2$ at time $t$, so that based on each point cloud $S_t$ and newly measured points $u_t$, we get $S_{t+1}$ with a procedure described in section \ref{sec:simulator} below.
We assume the a priori availability of a discriminator  (pre-segmenter) which for any point in $S$, decides whether the point is part of the Floor $F_t \subseteq S_t$ , Wall $W_t \subseteq S_t $ or any Target object $T_t \subseteq S_t$. In future work, this discriminator should be replaced with a network that automatically segments the point cloud into the desired classes /  parts. For instance, a PCT \cite{guo2021pct} discriminator network achieves an accuracy of $86.4\%$ and the state-of-the-art segmentation networks PointStack \cite{wijaya2022advanced} and SPoTr \cite{park2023self} achieve $87.2\%$ on the partShapeNet data set, which has $16$ object classes,  each  with different part segments.

We define a good path to be one where the agent finds a point cloud that covers the target objects as fast as possible. The cardinality of $T$, denoted by $|T|$, serves as a measure of success: the more points of the target objects have been found, the better we have mapped the objects. Our objective is therefore to find a sensor path $\mathbf{q} = \{q_0, q_1,...\}$ so that the target objects are found in as few steps as possible. As an example, this could be formalized with the objective function:
\begin{equation}
V(\mathbf{q}) = \sum_{t\geq 0} |T_{t+1} - T_t|
\end{equation}
although we will modify this for our RL algorithm, as explained later.

\subsection{Simulator}
\label{sec:simulator}
Based on the Robot Operating System (ROS) and the Gazebo simulator \cite{koenig2004design}, we develop a simulator for object search and coverage. Due to the fact that the sensor's movements are discrete, it is able to move only in discrete cells. This is why we will refer to these cells when discussing the environment's  dimensions. 

\begin{figure}
    \includegraphics[width=0.48\textwidth]{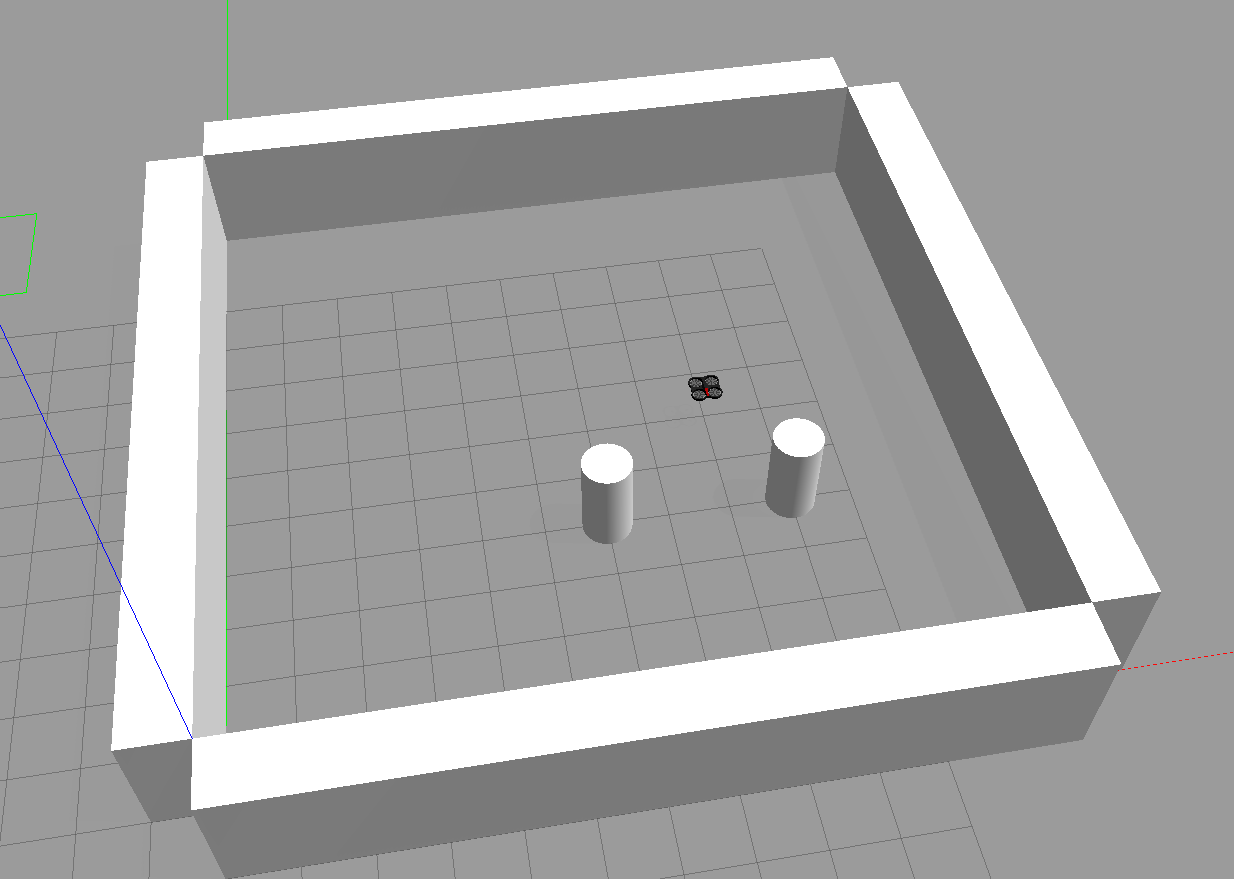}

    \includegraphics[width=0.48\textwidth]{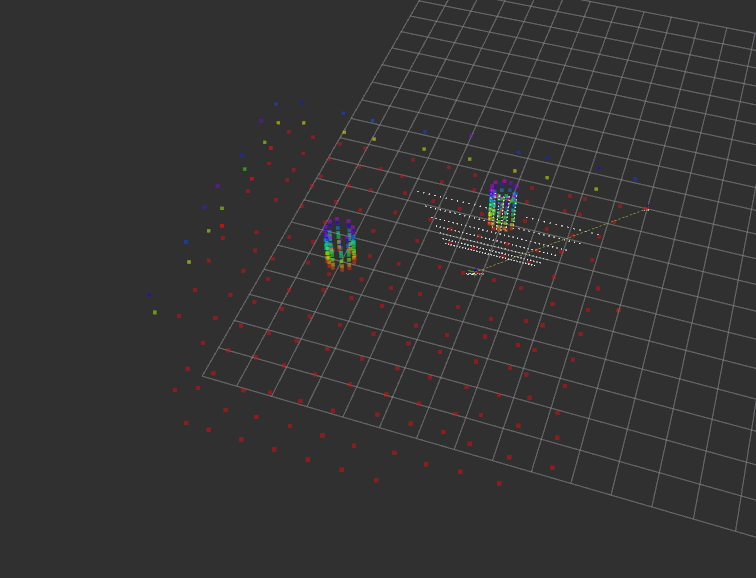} 
    \caption{Top: The environment with 2 cylinders, where our sensor is represented by the drone icon. Bottom: A point cloud of the environment with two cylinders placed in different locations than in the top figure.The white points represent the new incoming observation at time step $t$ of the agent, prior to its concatenation with the previous point cloud $S_t$.}\label{1_room_env}
\end{figure}

We conduct our tests in a one-room-scenario (see Figure \ref{1_room_env}). 
The environment has a size of $13 \times 13$ cells. Due to the walls, the agent's range of motion is restricted to $11 \times 11$ cells. Two cylinders with a diameter equal to 0.4 cells are uniformly randomly placed within the room's grid points, omitting the perimeter, such that the agent can always go around the cylinders.

 In our simulator, the point cloud for the 3D active target search and coverage task is measured by a simulated Kinect sensor with a range of $0.5-5$ cells. In order to accelerate the simulation, the resolution of the sensor is decreased to $640\cdot380$ points. This newly measured point cloud is concatenated with the previous measurements, after which a voxel filter \cite{zhou2018open3d} is employed to eliminate duplicates and to decrease the number of points, thereby reducing the training time and memory requirements.
We use two different voxel filters: one for the background (walls and floor) and one for the target objects. We utilized a voxel filter of $0.88$ cells for the background and $0.22$ cells for our target object. The reason is that while it is important to know which places the agent has already scanned, it is more important to identify the object (or its features) based on its shape, so we allow more points on the target object. The result of this measurement processing procedure is the new point cloud $S_{t+1}$.


\subsection{RL problem}
We model the problem as a partially observable Markov decision process, a tuple $\mathcal{M=	\bigl< S,A,O,R,T,U\bigr>}$ where $\mathcal{S}$ is the state space, $\mathcal{A}$ the action space, $\mathcal{O}$ the observation space, $\mathcal{R}$ is the reward function, $\mathcal{T}$ the transition dynamics and $\mathcal{U}$ generates the agent's observation. The state of the environment at time step $t$ is $s_t$, and the agent observes $o_t\sim\mathcal{U}(\cdot\mid s_t) \in \mathcal{O}$. The agent performs an action $a_t\in \mathcal{A}$ and receives a reward $r_t= R(s_t)$. The state of the environment changes to $s_{t+1}\sim \mathcal{T}(s_t,a_t)$. The objective is to maximize the return $G_t=\sum_{j=0}^T \gamma^j r_{t+j}$, where $\gamma \in (0,1)$ is a discount factor.

When following policy $\pi$, $\pi(a|s)$ is the probability that $a$ is executed in state $s$, where $a \in \mathcal{A}$ and $s\in \mathcal{S}$. The value of executing action $a$ in state $s$ under a policy $\pi$, $Q_{\pi}(s,a)$, is defined as the expected return $G_t$ when starting from $s$, performing action $a$, and then following policy $\pi$:
\begin{equation}
Q_{\pi}(s, a) := \mathbb{E}_{\pi}\left[G_{t} \mid s_{t}=s, a_{t}=a\right]
\end{equation}
Off-policy RL usually aims to learn the optimal value function 
$Q^*(s,a)= \max_{\pi} Q_{\pi}(s,a)$.

\textbf{State representation:} The state includes the true environment with its walls and targets, the gradually gathered point cloud with its three components (wall, floor, target), and the agent's pose.

\textbf{Action and transition:} The actions $\mathcal{A}$ are executed relative to the agent's coordinate system. To simplify the problem, the agent moves only one step forward, backward, left, right and rotates $90 ^{\circ}$ clockwise and  counterclockwise. It therefore stays in a horizontal place. An illegal transition $\mathcal{T}$ is defined as one which the agent moves into a wall or an object.

\textbf{Sensor and observation:}
The observation is the gradually gathered point cloud $S_t$ with its components $T_t$, $F_t$, $W_t$ as well as the pose of the agent $q_t$.

\textbf{Rewards:}
The agent's objective is to find the cylinders (targets) $T$ as quickly as possible; to help the agent explore more effectively, we additionally reward it for finding new floor points \cite{rosynski2022simulator}:  

\begin{equation}
r_t= |T_{t+1}|-|T_{t}|+|F_{t+1}|-|F_{t}|
\label{equ:reward}
\end{equation}
Owing to the larger number of points on the target object, the agent focuses its attention to scan the objects with priority.



\section{RL algorithm and network structure}

The experimental studies employ the Rainbow deep RL approach \cite{hessel2018rainbow} with spectral normalization \cite{gogianu2021spectral}. In addition, we employed collision-free exploration to enhance performance.  Every collision would cause the trajectory to end with a bad return, discouraging the agent from revisiting positions from which collisions are possible. In order to prevent this, we prohibit transitions that would result in the termination of an episode. Yet, prohibiting an action would never expose the agent to the collision penalties and allow the learn from them. Hence, we must store the illegal action with its associated reward and a termination flag in the experience replay buffer and take a different action for the remainder of the trajectory \cite{rosynski2022simulator}.

\begin{figure}
\includegraphics[page=1,width=.49\textwidth]{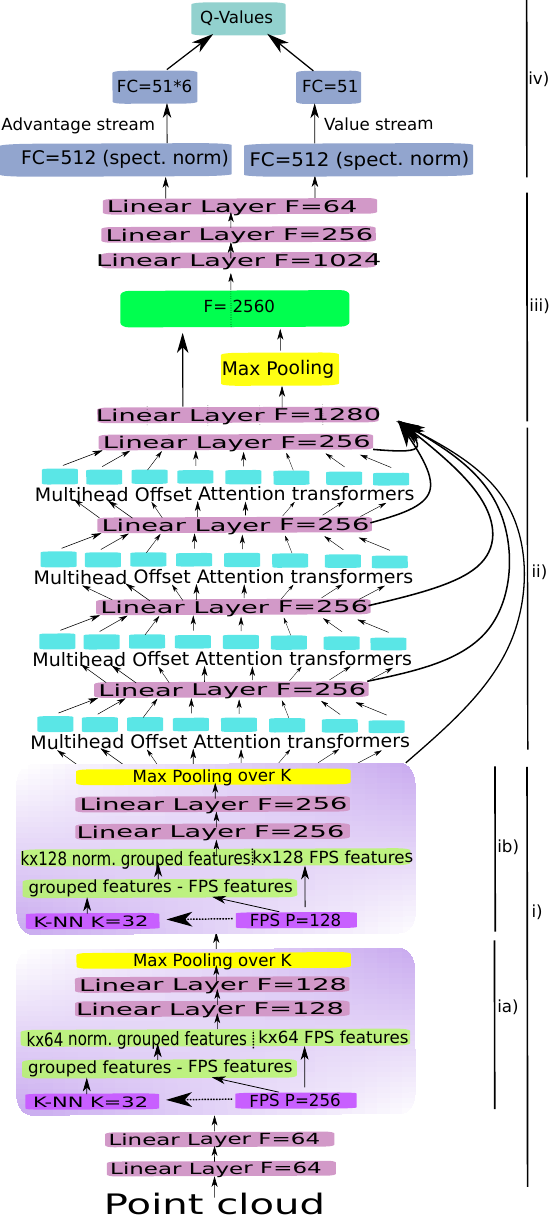}
\caption{The network architecture utilized for the experiment, where F denotes the feature vector's length and FC denotes a fully connected layer. Dashed lines represent concatenated layers. i) Two embedding blocks are used: ia) and ib). ii) The offset multihead attention block. iii) From PointNet, a component of the segmentation network was incorporated to preserve spatial information by performing max pooling on the points, copying them N times, and concatenating them with the original tensor. iv) A distributional and dueling network structure with spectral normalization.}
\label{fig:NN_structure}
\end{figure}
 
Next we will discuss the structure of the NN, which is depicted in Figure \ref{fig:NN_structure}, where the various components of the network are labeled. We begin with i) the embedding that PointNet++ \cite{qi2017pointnet++} introduced, then continue with ii) the multihead offset attention blocks. Next, iii) is a component that the PointNet segmentation network \cite{qi2017pointnet}  introduced to preserve spatial information.
The NN is completed by appending iv) the tail of the Rainbow structure with spectral normalization.

\begin{figure}[H]
    \center
\includegraphics[width=0.22\textwidth]{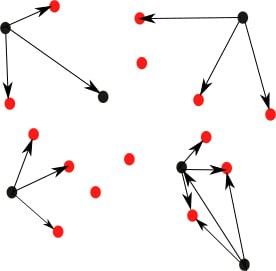}

\caption{The black points in this example of a 2D point cloud represent the sampled (FPS) points, which are the farthest points to each other. Every FPS point has three nearest neighbors. Owing to the uneven distribution of the points, one has no link whatsoever. }
\label{fig:FPS}
\end{figure}

\textbf{i) Embedding.} The network embeds the points using deep hierarchical feature learning, introduced by PointNet \cite{qi2017pointnet++}. Given a point cloud, the key idea is that a farthest point sampling (FPS) algorithm seeks out representative points on which a K-nearest neighbors algorithm is then applied, see Figure \ref{fig:FPS}. 

The FPS technique is also used in the PCT network. The segmentation PCT network embedding block differs from the classification PCT embedding block in that for segmentation, every point must be kept because it needs to be assigned to a class by the output of the NN. This indicates that FPS is effectively skipped and only the K-nearest neighbor algorithm is applied to each point; in other words, the FPS input point cloud is identical to the output point cloud (all points are selected). 

We demonstrate that it is possible to reduce the size of the network by embedding neighbor information in the feature vectors of certain  points chosen by the FPS algorithm. 
This reduces the parameter space of the network as we explain below, and also achieves better results (see Figure  \ref{fig:5k_expNA1}). The downside is that the   features of the FPS points must learn the relationship between the points ${P_{\mathrm{fps}}} \in \mathcal{R}^{B\times S\times F}$ selected by FPS and their neighbors $P_{\mathrm{kk}} \in \mathcal{R}^{B\times S\times K\times F}$ in the embedding blocks, which requires additional training steps. Here, $B$ is the batch size, $S$ the number of sampled points by FPS, $K$ the number of neighbors, and $F$ the the length of the feature vector.

The features of each FPS point's \textit{K} nearest neighbors are subtracted from the FPS features to normalize the local features, and the result is called grouped features (see the embedding block in Figure \ref{fig:NN_structure}). Next the FPS features of the selected points (with $K$ instances) and the grouped features are concatenated. After applying two linear layers we apply max pooling over $K$. The method is repeated using the new $P_{\mathrm{fps}}$ points,  which are used as inputs for a new embedding block (see again Figure \ref{fig:NN_structure}). This results in a hierarchical organization. 

To accurately estimate the state value in DRL, all points must be represented in the features of the FPS points $P_{\mathrm{fps}}^{b}$ in each embedding block $b$. Figure \ref{fig:FPS}  depicts an example of a point that would not be represented in the feature vector. Because the points in our experiments are not distributed evenly, we must carefully select the number of neighbors and samples to ensure that every point $P^b$ in the embedding block $b$ is represented in the set of of sampled FPS points $P_{\mathrm{fps}}^b$. This implies that the number of input points $P_b$ in the embedding block $b$ must be less than the product of the number of FPS points and the number of neighbors: 
\begin{equation}
|P^b|<|P_{\mathrm{fps}}^b| \cdot |P_{\mathrm{knn}}^b|
\label{equ:min}
\end{equation}

\textbf{ii) Multihead attention.} In transformers or Attention Networks, the correlation between two distinct weight matrices is calculated and evaluated by a third matrix.  In the original transformer paper \cite{vaswani2017attention} that introduced multi-head attention, it was stated that by implementing multiple heads, distinct features are learned by each head.
In contrast, each transformer block in PCT \cite{guo2021pct} has only one head.  To investigate whether a multi-head attention block enhances the outcomes of point-cloud-based mapping in RL, we created one with the offset attention structure introduced in the PCT network  \cite{guo2021pct}. Moreover, to combine the heads, we concatenated them and added an extra linear layer which reduces the output of the features back to $256$.

\textbf{iii) Max pooling.}  Max pooling is applied over the points. To preserve spatial information, we use the technique from PointNet \cite{qi2017pointnet}, i.e. we concatenate $N$ copies of the result of the max pooling layer with the previous layer, where $N$ is the number of points. 

\textbf{iv) RL tail.} On the network's tail, a dueling decomposition and a distributional representation \cite{bellemare2017distributional} are applied, both of which are components of the Rainbow network \cite{hessel2018rainbow}. The distributional part, instead of directly representing expected values of the returns, represents the distribution of returns over a discrete number of bins defined over a finite interval from $V_{min}$ to $V_{max}$. \cite{bellemare2017distributional}. The dueling network structure divides the network into an advantage and a value stream. A naive approach would be to add up the two streams to get $Q(s,a)= V(s) + A(s,a)$, but the problem is that $V$ and $A$ would be unidentifiable in the sense that given $Q$ we cannot recover $V$ and $A$ uniquely. To solve this issue we use $Q(s,a)=V(s)+\left( A(s,a)-\frac{1}{|A|}\sum_{a'}A(s,a')\right)$. In addition, spectral normalization is added to the first fully connected layer of the value and advantage stream.

\section{Experiments}
\subsection{Experiment setup}
We conducted two separate batches of experiments. In one, the goal is to improve the network's architecture. In the other, we train the best network found in the first set of experiments to compete with a greedy baseline strategy.

\textbf{Network architecture.}
We train our agents over $5,000$ episodes consisting of $65$ steps each. Because our agent employs an $\epsilon$-greedy exploration strategy \cite{sutton2018reinforcement}, we implemented a linear decay function for exploration beginning at  $\epsilon=1$ (full exploration) and completing exploration  $(\epsilon=0$) after $200,000$ steps. Training began after $15,000$ steps. Unless otherwise specified, the number of neighbors is $32$ in all experiments. The maximum number of points the agent could find was $512$. For the $C51$ distributional representation \cite{bellemare2017distributional} we define the minimum $V_{\mathrm{min}}=0$ and maximum $V_{\mathrm{max}}=415$ of the returns, which we experimentally established. Other hyper-parameters are taken from \cite{hessel2018rainbow}.

In our network experiments, we first examine how FPS affects training in terms of performance, before comparing the various network structures.
Secondly, we evaluate whether and how multi-head attention influenced learning.
In our final ablation, we examine how strongly FPS with neighbor embedding influences the learning, by removing components ia) and ib) of Figure \ref{fig:NN_structure}).


\textbf{Baseline comparison.}
We train our agent for $20,000$ episodes, each of which consists of $65$ steps. We again utilized a linear decay function for exploration, with the agent completing its exploration after $200,000$ steps. For the evaluation of the finale learned policy, we perform $100$ steps over $100$ episodes with each agent. The maximum number of points the agent could find was $512$. 

As our baseline, we selected a greedy algorithm, which simulates all potential actions and selects the one with the highest immediate reward (\ref{equ:reward}). Because it is simulating based on the ground truth, it has an information advantage: it knows the observation prior to scanning, which means it would not be implementable in practice.
Due to the lack of research in the field of active target search and coverage problems, we make the decision to utilize a greedy algorithm as a baseline, with the same reward function as RL. The alternative would be to employ a state-of-the-art active coverage algorithm \cite{guedon2022scone, guedon2023macarons} that focuses on the coverage of the entire environment rather than only of the target objects, --  a different objective.

\subsection{Results \& discussion}
As explained above, we will first describe the results with various network configurations, and then we will contrast our approach with the greedy baseline. We encode the various architectures choices as the following string: $NsXXXsXXXhX$, where $N=C$ stands for classification and $N=S$ for segmentation, $sXXX$ for FPS sampling with $XXX$ the number of FPS samples per embedding block, and $hX$ for the number of transformer heads. 

\textbf{Network architecture.}
We have experimentally investigated the following network configurations.
First, for one head, we compare the classification network structure $(Cs256s128h1)$, with the segmentation embedding block from the PCT network $(Ss512s512h1)$. In the second network, the number of FPS points is $512$ in both embedding blocks in order to preserve all points for the higher layers. We label this experiment NA1. Next, we examine a network with eight heads $(Cs256s128h8)$ (see Figure  \ref{fig:NN_structure}) with the segmentation embedding block also with 8 heads $(Ss256s128h8)$, in experiment NA2. Next, in NA3, we compare a network with a single embedding block containing $32$ FPS samples with one transformer head $(Cs32h1)$ versus eight transformer heads $(Cs32h8)$ to investigate, on the one hand, the influence of the number of heads when the network's parameters are reduced and, on the other hand, whether or not the network can learn when the number of points is reduced from $512$ to $32$ points by FPS. Last but not least, in NA4, we train our best agent fully and compare it to an agent ($Ss0s0h8$) that omits the components ia) and ib) of Figure \ref{fig:NN_structure}, to determine the effect of FPS with neighbor embedding.

\begin{figure}
\center
    \includegraphics[width=0.4\textwidth]{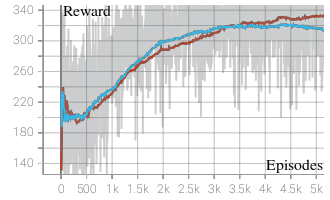} 
    \caption{NA1. The $Ss512s512h1$ network is depicted in blue, while  $Cs256s128h1$ is depicted in red. To smooth the graphs we plotted $x$, computed with the formula: $x_{\tau+1} = 0.99x_{\tau} + (1 - 0.99) R_{\tau}$, where $R_{\tau}$ is the return of episode $\tau$.}
    \label{fig:5k_expNA1}
\end{figure}

Experiment NA1 (Figure \ref{fig:5k_expNA1}) shows that the segmentation network $Ss512s512h1$ is learning faster than classification network $Cs256s128h1$, but the classification network exceeds the segmentation network at the end. The reason can be that $Ss512s512h1$ is overfitting due to more parameters than in $Cs256s128h1$.

\begin{figure}
    \includegraphics[width=0.4\textwidth]{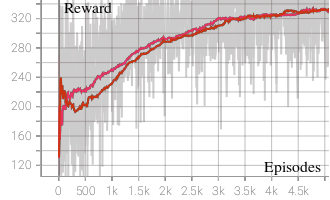} 
    \caption{NA2. The  $Cs256s128h1$ network is depicted in red vs. the $Cs256s128h8$ in pink. To smooth the graphs we used the same formula as before.}
    \label{fig:5k_expNA2}
\end{figure}

\begin{figure}
    \includegraphics[width=0.4\textwidth]{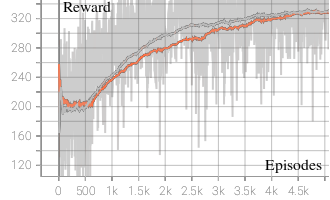} 
    \caption{NA3. Orange depicts the $Cs32h1$ network with a single embedding block and with 32 FPS samples, and grey depicts the $Cs32h8$ network. To smooth the graphs we plotted $x$, as in Figure \ref{fig:5k_expNA2}.}
    \label{fig:5k_expNA3}
\end{figure}

In NA2  (Figure \ref{fig:5k_expNA2}) the network $Cs256s128h8$ with 8 heads is learning faster than $Cs256s128h1$, but then converges to the same result.
In NA3  (Figure \ref{fig:5k_expNA3}), we observed not only that the networks were able to learn but also that the network $Cs32h8$ with 8 heads again learned faster and converged to the same result. We observed this behavior  in all our experiments\footnote{We also attempted the multi-head attention on a classification task where we used the standard PCT classification network from \cite{guo2021pct} where we changed to 8 attention heads and obtained the same results, namely, that the classification network learned faster than $Cs32h1$ but converged to the same outcomes. These classification and the other network configuration results are not given due to lack of space.}: learning a multi-head attention is faster, but leads to the same result. One possible explanation is that all transformer heads are actually learning the same features, unlike neural networks for language processing \cite{vaswani2017attention}. However, because we merge the output with a linear layer of the same size, it is possible that these heads are averaging the poorly trained features and achieving more stable results at the beginning of the training. 

\begin{figure}
    \centering
    \includegraphics[width=0.4\textwidth]{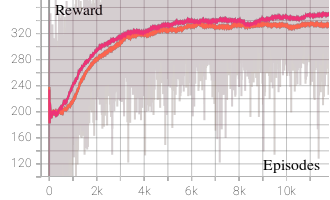}
    \caption{NA4. The  $Cs256s128h8$ network depicted in red vs. $Ss0s0h8$ in orange. To smooth the graphs we plotted $x$, computed with the formula: $x_{\tau+1} = 0.997x_{\tau} + (1 - 0.997) R_{\tau}$, where $R_{\tau}$ is the return of episode $\tau$.}
\end{figure}
\begin{figure}
    \centering
    \includegraphics[width=0.5\textwidth]{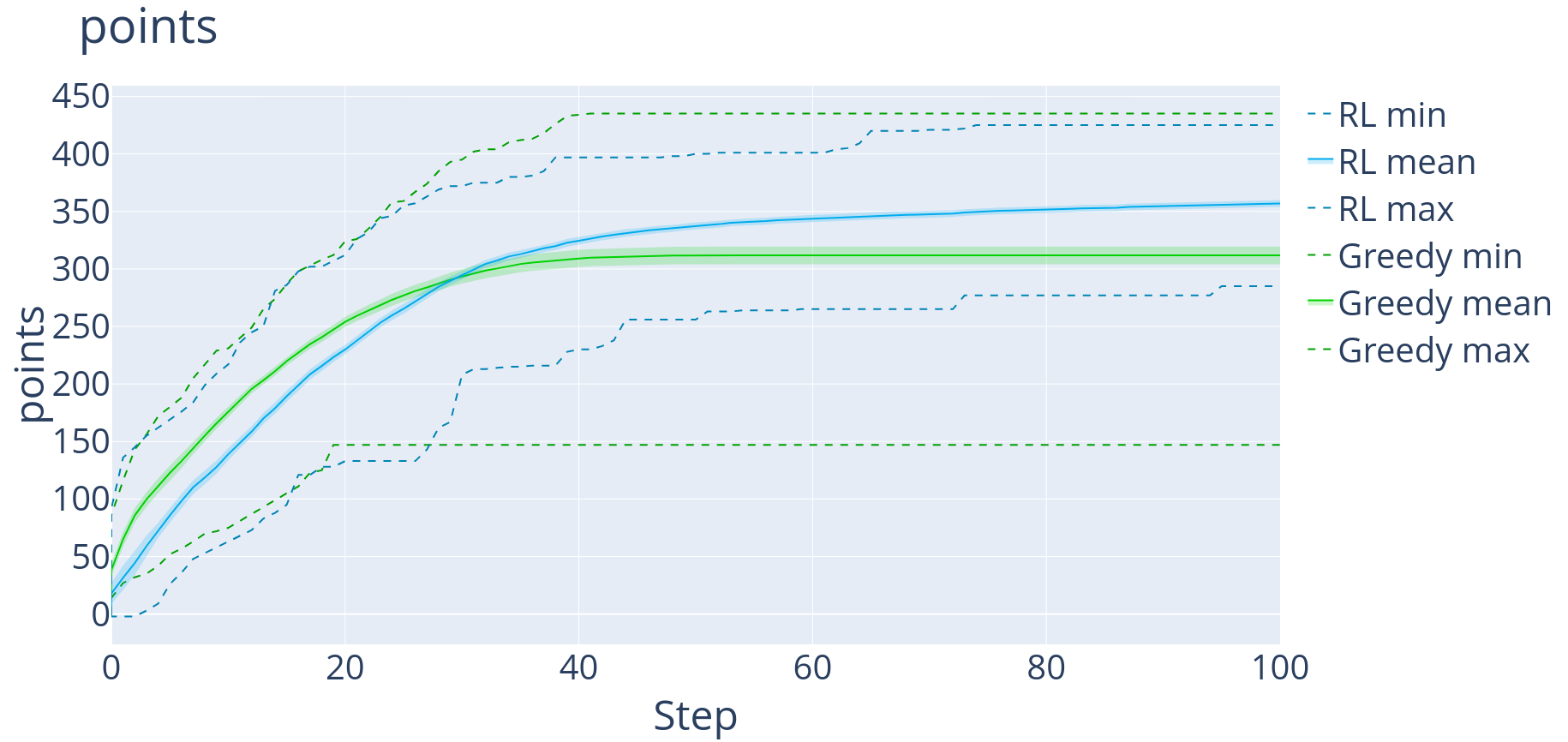} 
    \caption{The graph depicts 100 steps over 100 episodes. Green shows the greedy agent, which converges after 100 steps to $311.8$ points on average. Blue shows the RL agent ($Cs256s128h8$), which reaches $356.9$ points.  }\label{fig:baselineEx}
\end{figure}


Experiment NA4 in Figure \ref{fig:baselineEx} shows that the KNN-embedding has a positive influence on the agents performance.

\textbf{Baseline comparison:}
Figure \ref{fig:baselineEx} shows our RL versus greedy baseline experiment. In green, the greedy baseline is given, the best RL agent, $Cs256s128h8$ is shown in blue. The dashed lines indicate the best / worst trajectory over the span of 100 episodes. The pale blue / green color indicates the agent's $95\%$ confidence on the mean.
The experiments indicate that the RL agent outperforms the greedy algorithm, which tends to become stuck after approximately 45 steps. Moreover, the confidence interval of the RL agent is tighter, and its worst trajectory is far better than that of the greedy algorithm. Hence, the RL agent is more reliable than the greedy one.

\begin{figure*}
\center
    \includegraphics[width=0.3\textwidth]{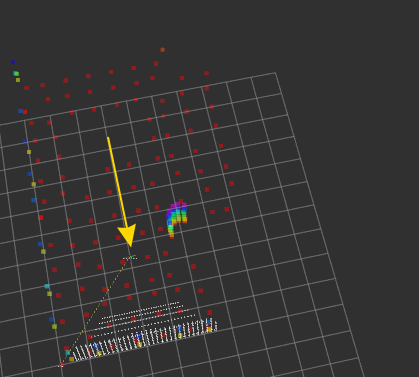} \vspace{0.05cm}
    \includegraphics[width=0.3\textwidth]{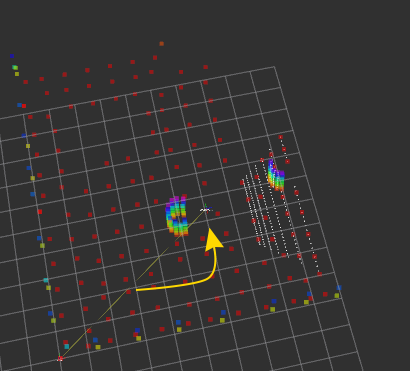} 
    \includegraphics[width=0.3\textwidth]{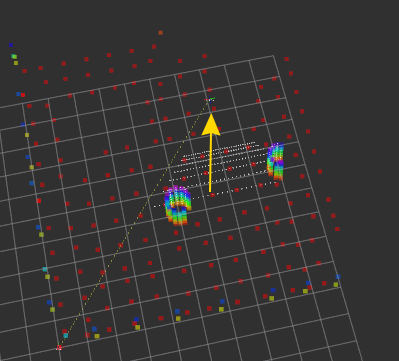} 
    \includegraphics[width=0.3\textwidth]{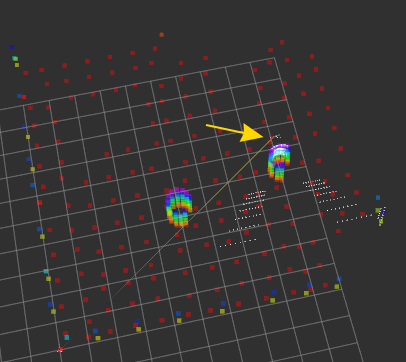} 
    \includegraphics[width=0.3\textwidth]{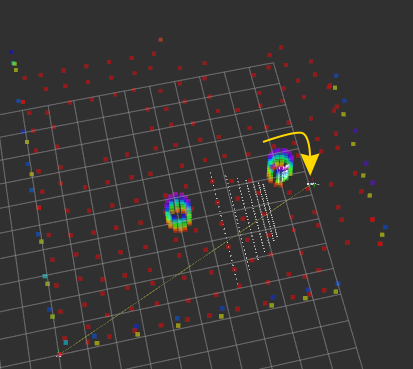}
    \includegraphics[width=0.3\textwidth]{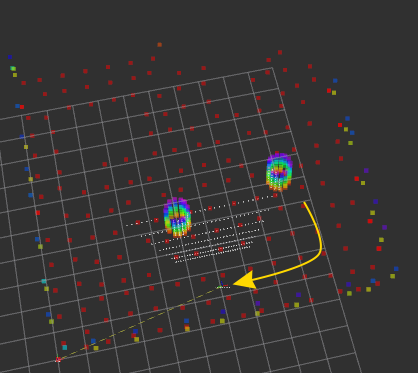} \caption{A path of a fully trained agent. Top row from left: 1) The episode begins and after some steps the first cylinder is discovered. 2) The agent moves around and discovers the second cylinder. 3) The agent moves forward and turns around to scan both cylinders from another perspective. Bottom row from the left: 4) The agent starts to move closer to the second cylinder. 5) The agent moves around the second cylinder and scans it. 6) The agent moves forward and scans the first cylinder from a different perspective.}\label{fig:trajectory}
\end{figure*}

Figure \ref{fig:trajectory} depicts the agent's trajectory. We can see how the agent learned to concentrate on and navigate around the cylinders.

\section{Conclusion}
In this study we demonstrated that an RL agent can discover an efficient path for a mobile 3D sensor to locate and cover target objects in the scene. We experimented with various configurations of the neural network and investigated the applicability of a multi-head attention for point clouds, concluding that it can obtain a slight speed advantage in learning. Finally, and most importantly, the RL method worked better than a greedy baseline.

Future research will examine whether RL is effective in environments with more rooms and objects, including clutter objects that do not need to be covered. Every extra object and room increases the required number of points, which not only increases the computation and complexity but also the training time. It should also be investigated whether it is possible to train the RL agent without the a priori availability of a discriminator. 




\bibliographystyle{ieeetr}
\bibliography{source} 

%
%


\end{document}